\documentclass[journal]{IEEEtran}
\IEEEoverridecommandlockouts
\usepackage{cite}
\usepackage{amsmath,amssymb,amsfonts}
\usepackage{amsthm} 

\usepackage{algorithm}      
\usepackage{algpseudocode} 
\usepackage{algorithmicx}  
\usepackage{makecell}
\usepackage{graphicx}
\usepackage{textcomp}
\usepackage{amsfonts} 
\usepackage[colorlinks, bookmarksopen, bookmarksnumbered, citecolor=blue, linkcolor=blue, urlcolor=blue]{hyperref}
\usepackage{xcolor}     
\usepackage{balance}    
\usepackage{booktabs}   
\usepackage{amsthm}     
\usepackage{float}      
\usepackage{subfig}
\usepackage[capitalize]{cleveref}
\usepackage{caption}
\let\labelindent\relax
\usepackage{enumitem}
\newtheorem{theorem}{Theorem}

\newtheorem{lemma}{Lemma}

\begin{document}

\title{\vspace{-0.3cm}Uplink Rate Maximization for Pinching Antenna- Assisted Covert Backscatter Communication
}

\author{Yulei Wang, 
        Yalin Liu, 
        Yaru Fu, 
        Yuanwei Liu,~\IEEEmembership{Fellow,~IEEE}
\thanks{Y. Wang is with the 
College of Electronics and Information Engineering, South-Central Minzu University. 
Y. Liu, Y. Fu are with the School of Science and Technology, Hong Kong Metropolitan University, Hong Kong.
Y. Liu is with the Department of Electrical and Electronic Engineering, The University of Hong Kong, Hong Kong.
}

\vspace{-0.8cm}}

\markboth{Journal of \LaTeX\ Class Files,~Vol.~14, No.~8, August~2025}%
{Shell \MakeLowercase{\textit{et al.}}: Bare Demo of IEEEtran.cls for IEEE Journals}


\maketitle

\begin{abstract}
The emerging pinching antenna (PA) technology enables flexible antenna positioning for creating line-of-sight (LoS) links, thus offering substantial potential to facilitate ambient signal-based backscatter communication (BSC). This paper investigates PA-assisted BSC for enhanced communication and covertness in the presence of a randomly distributed eavesdropper. 
An optimization problem is formulated to maximize the uplink covert transmission rate by jointly optimizing the transmit power and antenna positions while satisfying both communication reliability and covertness constraints. An alternative optimization (AO)-based framework is proposed to solve this problem. Numerical results demonstrate that the proposed PA-BSC effectively mitigates the double near-far problem, where energy harvesting and backscatter transmission degrade simultaneously due to distance disparities, thereby improving downlink energy harvesting and uplink data transmission while maintaining covertness performance under practical deployment scenarios.
\end{abstract}
\vspace{-0.3cm}

\IEEEpeerreviewmaketitle

\vspace{-0.3cm}
\section{Introduction}
\begin{figure}[t]
    \centering
    \includegraphics[scale=0.5]{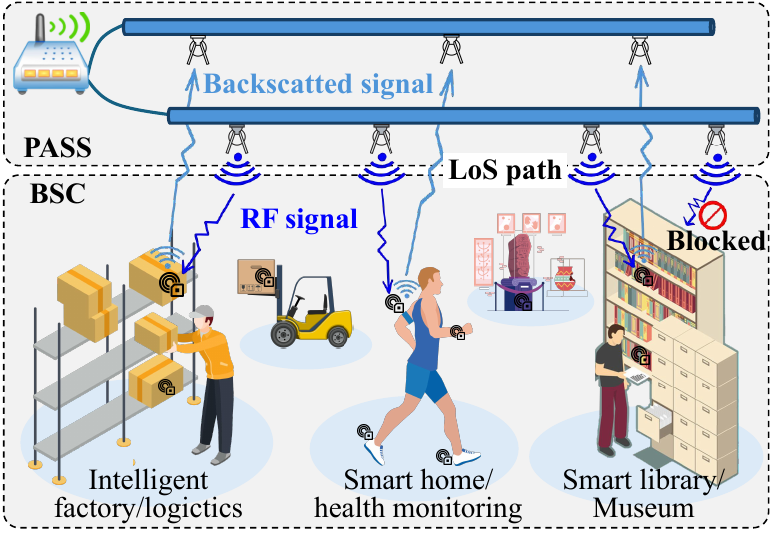}
    \caption{PA-assisted backscatter communication scenarios.}
    \label{Fig-PBSC-Scenario}
\end{figure}

\IEEEPARstart{B}{ackscatter} communication (BSC)~\cite{wang24IoTJ}, which leverages ambient radio-frequency (RF) signals to modulate and transmit data, has garnered significant attention as a promising solution for low-power, energy-efficient wireless communication systems in sixth-generation (6G) networks. As illustrated in~\cref{Fig-PBSC-Scenario}, backscatter devices (BDs) are increasingly deployed in diverse applications, including intelligent factories, logistics, smart homes, healthcare, smart libraries and museums~\cite{zhao23IoB}. In such system, BDs harvest energy from ambient RF signals, modulate the signals, and reflect them to convey information. However, BDs often suffer from the ``\textit{double near-far}" problem~\cite{papani25Power, fu25NOMA}, wherein both energy harvesting and data transmission deteriorate due to distance variations between the RF emitter, the BD, and the reader. This issue severely constrains the performance and scalability of BSC networks, hindering their broader application.

Pinching antenna (PA) technology, first demonstrated by NTT DOCOMO in 2021~\cite{suzuki22pinching}, provides an innovative solution to the challenges associated with \textit{double transmission} between transceivers in wireless systems~\cite{fu25NOMA, ding25flexible}. By enabling flexible positional adjustments, PAs can establish and maintain line-of-sight (LoS) links, thereby reducing signal attenuation and improving transmission efficiency, especially at high frequency bands, such as millimeter-wave (mmWave) and terahertz (THz)~\cite{wang25Indoor}. Unlike conventional transceivers with fixed positions, the transmit PA (TPA) and the receive PA (RPA) in pinching antenna systems (PASS) are spatially decoupled and can be optimally positioned to enhance both downlink RF signals transmission and uplink data communication. This adaptability makes PASS highly promising and applicable for enhancing BSC performance.

Existing studies~\cite{papani25Power, Wang25Power, peng25PoweredMEC} have investigated energy harvesting and data transmission in PA-assisted wireless powered networks, where users typically employ a harvest-then-transmit protocol, harvesting energy in the downlink and transmitting data in the uplink in a time-division manner. However, such a model does not apply to PA-BSC networks, where BDs simultaneously harvest energy from the TPA (acting as an RF emitter) and backscatter information to the RPA (acting as a tag reader). Furthermore, due to the broadcast nature of wireless signals and weak transmit power of BD, backscatter transmissions are inherently vulnerable to eavesdropping and malicious interception. Although recent works~\cite{jiang25Covert, sun25PLS} have explored covert communication in PASS, the joint guarantee of reliability and covertness in PA-BSC networks remains an open problem. This critical gap motivates our work.

To the best of our knowledge, this paper is the first to investigate the uplink covert transmission rate maximization in PA-BSC networks. Our contributions are summarized as follows:
\textbf{First}, we propose a three-dimensional (3D) spatial model for a PA-BSC network, wherein a BD harvests energy from a TPA and backscatters data to an RPA, in the presence of a randomly located eavesdropper (Eve). The TPA and RPA are flexibly adjustable along two ceiling-mounted waveguides.
\textbf{Second}, we develop a practical covert transmission model that accounts for Eve’s detection performance under both noise and location uncertainties.
\textbf{Third}, we formulate an optimization problem to maximize the uplink covert transmission rate and propose an alternating optimization method to jointly design the transmit power and antenna positions under reliability and covertness constraints.
\textbf{Finally}, numerical results demonstrate the effectiveness of our algorithm, indicating that proposed system mitigates the ``\textit{double near-far}" effect, where energy harvesting and backscatter transmission degrade simultaneously due to distance disparities, meanwhile improves the transmission performance compared to baseline scheme. 
\textit{This work highlights the potential of PASS as a key enabler for future energy-efficient and secure BSC systems.}

\vspace{-0.3cm}
\section{System Model}\label{Sec-Sys}
We consider a PA-BSC network deployed in an indoor service room, e.g., a factory or exhibition hall, as illustrated in~\cref{Fig-PBSC-Overview}. The network comprises an access point (AP), a BD, and an unauthorized Eve. The AP connects two lossless dielectric waveguides via feed points: one equipped with a TPA (for transmitting signals to BD) and the other with an RPA (for receiving signals from BD). Meanwhile, Eve attempts to detect the information exchange between the AP and the BD.
\vspace{-0.3cm}
\subsection{Network Topology}
\begin{figure}[t]
    \centering
    \includegraphics[scale=0.46]{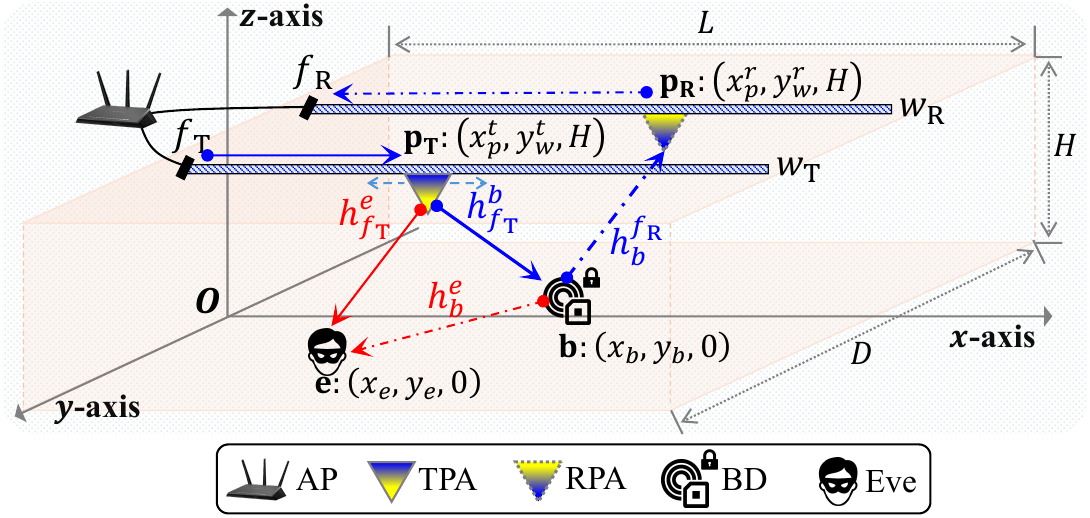}
    \caption{A typical PA-BSC network with an eavesdropper.}
    \label{Fig-PBSC-Overview}
\end{figure}

The service room, denoted by $\mathcal{A}$, is modeled as a rectangular cuboid with length $L$, width $D$, and height $H$. Two parallel waveguides with same length $L$ are installed along the length direction at the ceiling. Let $p_{\mathrm{T}}, p_{\mathrm{R}}, w_{\mathrm{T}}, w_{\mathrm{R}}, f_{\mathrm{T}}$, and $f_{\mathrm{R}}$ denote the TPA, the RPA, their waveguides and the feed points, respectively. The positions of $p_{\mathrm{T}}$ and $p_{\mathrm{R}}$ can be flexibly shifted along $w_{\mathrm{T}}$ and $w_{\mathrm{R}}$, respectively. The BD and Eve, denoted by $b$ and $e$, are assumed to be randomly located on the floor. To model the locations of all nodes in $\mathcal{A}$, a 3D Cartesian coordinate system is established with the origin $\mathcal{O}$ located at the center of the width side on the floor (the same side where the feed points are placed), as shown in~\cref{Fig-PBSC-Overview}. Hence, the coordinates of $p_{\mathrm{T}}$, $f_{\mathrm{T}}$, $p_{\mathrm{R}}$, and $f_{\mathrm{R}}$ can be represented as
$\mathbf{p_{\mathrm{T}}}:(x_p^t, y_w^t, H)$,~
$\mathbf{f_{\mathrm{T}}}:(0, y_w^t, H)$,
$\mathbf{p_{\mathrm{R}}}:(x_p^r, y_w^r, H)$,~
$\mathbf{f_{\mathrm{R}}}:(0,y_w^r,H)$, where $\{x_p^t, x_p^r\} \in [0,L]$. The coordinates of $b$ and $e$ are represented as $\mathbf{b}:(x_b, y_b, 0)$, $\mathbf{e}:(x_e, y_e, 0)$ with $\{x_b, x_e\} \in [0,L]$ and $\{y_b, y_e\} \in [-D/2, D/2]$.

\vspace{-5pt}
\subsection{Channel Model}\label{Subsec-PA}
The AP injects RF signals into the transmit waveguide $w_{\mathrm{T}}$. These signals propagate from $f_{\mathrm{T}}$ to $p_{\mathrm{T}}$, and are then broadcast into space. The overall channel is affected by phase shifts within the waveguide, free-space phase shifts, and free-space path loss. Let $h_{f_\mathrm{T}}^{i}$ denote the channel coefficient from $f_{\mathrm{T}}$ to a receiver $i\in{b,e}$. It can be formulated as follows:
\begin{equation}
\small
    h_{f_\text{T}}^i = \underbrace{e^{-j\frac{2\pi}{\lambda_g} d_{f_\text{T}}^{p_\text{T}}} }_{\text{in-waveguide phase shift}} \cdot \underbrace{e^{-j\frac{2\pi}{\lambda} d_{p_\text{T}}^i} }_{\text{free-space phase shift}} \cdot \underbrace{\eta^{\frac{1}{2}}\left(d_{p_\text{T}}^i\right)^{-1}}_{\text{free-space path loss}}, 
\end{equation}
where $\lambda_g = \frac{\lambda}{n_{\text{eff}}}$ denotes the guided wavelength with $n_{\text{eff}}$ being the effective refractive index of the dielectric waveguide, $\lambda$ is the signal wavelength, $\eta = \lambda^2/16\pi^2$ is the reference path loss at $1$m, $d_{f_\text{T}}^{p_\text{T}}=\|\mathbf{f_\text{T}} - \mathbf{p_\text{T}}\|$, $d_{{p_\text{T}}}^{i}=\|\mathbf{p_\text{T}} - \mathbf{i}\|$ with $\mathbf{i} \in \{\mathbf{b}, \mathbf{e}\}$. Upon detecting the RF signals, the BD modulates and reflects its data toward $p_{\mathrm{R}}$, which then travels to $f_{\mathrm{R}}$ through $w_{\mathrm{R}}$. The channel from $b$ to $f_{\mathrm{R}}$ is
\begin{equation}
\small
    h_b^{f_\text{R}} = 
    \underbrace{\eta^{\frac{1}{2}}\left(d_b^{p_\text{R}}\right)^{-1}}_{\text{free-space path loss}}
    \cdot \underbrace{e^{-j\frac{2\pi}{\lambda} d_b^{p_\text{R}}} }_{\text{free-space phase shift}}
    \cdot \underbrace{e^{-j\frac{2\pi}{\lambda_g} d_{p_\text{R}}^{f_\text{R}}}}_{\text{in-waveguide phase shift}}, 
\end{equation}
where $d_b^{p_\text{R}}=\|\mathbf{b}- \mathbf{p_\text{R}}\|$ and $d_{p_\text{R}}^{f_\text{R}}=\|\mathbf{p_\text{R}} -\mathbf{f_\text{R}}\|$.

Since both the BD and Eve are located on the ground, the channel between them experiences distance-dependent path loss and multipath fading. Hence, the BD-to-Eve channel $h_b^e$ is modeled as $h_b^e = l_b^e g_b^e$, where $l_b^e = \sqrt{\eta (d_b^e)^{-\alpha}}$ captures the large-scale fading with path loss exponent $\alpha$, $g_b^e$ models the small-scale fading, and $d_b^e = \|\mathbf{b} - \mathbf{e}\|$. 

\vspace{-0.3cm}
\subsection{Transmission Model}
The AP injects RF signals to the TPA, which then broadcasts to the BD. 
Let $y_b$ denote BD's received signal from the AP, which can be written as $y_b = \sqrt{P_0} h_{f_\text{T}}^b s_0$, where $P_0$ is the transmit power of $p_{\mathrm{T}}$ and $s_0$ is the normalized transmitted signal with $\mathbb{E}[|s_0|^2]=1$. Since the BD comprises only passive components, the thermal noise at $b$ is negligible. Thus, the harvested power at $b$, denoted by $P_b$, can be calculated as follows: 
\begin{equation}
\begin{aligned}
\small
& P_b = P_0 \left|h_{f_\text{T}}^b\right|^2 
= P_0 \eta \left(d_{p_\text{T}}^b \right)^{-2}.
\end{aligned}
\label{Eq-phase2}
\end{equation}
After receiving the RF signals, the BD harvests a portion of energy from the signals to reflect its data to the RPA and propagates it to the AP. Let $\kappa$ be the power fraction of the BD for reflecting,
the remaining fraction $1-\kappa$ is reserved for energy harvesting. Let $y_p$ denote the received signal at $p_{\mathrm{R}}$, which is given by $y_p = \sqrt{\zeta\kappa P_b}\, h_b^{f_\text{R}} s_b s_0 + W_p$, where $\zeta \in[0,1]$ denotes the backscattering efficiency~\cite{lu17wireless}, 
$s_b$ is the BD’s modulated signal with $\mathbb{E}[|s_b|^2]=1$, and $W_p\sim \mathcal{CN}(0, \sigma_p^2)$ is the additive white Gaussian noise (AWGN) at $p_{\mathrm{R}}$. Therefore, the received signal power at $p_{\mathrm{R}}$ can be calculated as follows:
\begin{equation}
\begin{aligned}
& P_a = \zeta \kappa P_b | h_b^{f_\text{R}}|^2
= \zeta \kappa P_0 \eta^2 \left(d_{p_\text{T}}^b d_b^{p_\text{R}} \right)^{-2}.
\end{aligned}
\label{Eq-Pp}
\end{equation}
Thus, the overall signal-to-noise ratio (SNR) from the BD to the AP can be calculated by $\gamma_a =P_a/\sigma_p^2$ and the achievable transmission rate $R_b$ of $b$ can be calculated by $R_b = B \log_2\!\left(1+\gamma_a\right)$, where $B$ is the allocated bandwidth.

Eve detects the RF signals from both the TPA and the BD. Upon receiving the RF signals, Eve performs a binary hypothesis test to determine whether $b$ is actively backscattering. The received signals at $e$ can be expressed as follows~\cite{wu2025covert}:
\begin{equation}
\small
y_e = \begin{cases}\sqrt{P_0} h_{f_\text{T}}^e s_0 + W_e, &\mathcal{H}_0, \\ 
\sqrt{\zeta\kappa P_b} h_b^e s_b s_0 + \sqrt{P_0} h_{f_\text{T}}^e s_0 + W_e, &\mathcal{H}_{1}, \end{cases}
\label{Eq-Eve-Siganl}
\end{equation}
where $\mathcal{H}_0$ denotes the null hypothesis that $b$ is inactive, and $\mathcal{H}_1$ denotes the alternative one that $b$ is transmitting to $p_{\mathrm{R}}$, $W_e \sim \mathcal{CN}\left(0, \sigma_e^2\right)$ represents the AWGN at $e$. Under the presence of Eve, we will model the security of PA-BSC by using the signal detection performance of Eve, which is given in~\cref{Sec-Dec}. 
\vspace{-0.3cm}

\section{Detection Performance}\label{Sec-Dec}
This section first defines the signal detection performance of Eve in terms of the optimal detection threshold and the detection error probability (DEP). To model the security of PA-BSC,  the lower bound of Eve's DEP is derived.

\vspace{-0.3cm}
\subsection{Detection Performance At Eve}
The received power $P_e$ at $e$ is given by $P_e \triangleq |y_e|^2 \mathop{\gtrless}_{\mathcal{D}_0}^{\mathcal{D}_1} \Gamma_{\text{th}}$,
where $\Gamma_{\text{th}}$ is the energy detection threshold at $e$. The terms $\mathcal{D}_0$ and $\mathcal{D}_1$ represent the binary decisions of whether $b$ is transmitting to $p_{\mathrm{R}}$, where $\mathcal{D}_0$ supports $\mathcal{H}_0$, and $\mathcal{D}_1$ supports $\mathcal{H}_1$. 
Substituting the formula of $y_e$, $P_e$ can be calculated by
\begin{equation}
\small
P_e = |y_e|^2= \begin{cases}  \Delta_1+ \sigma_e^2, & \mathcal{H}_0, \\ \Delta_2 + \sigma_e^2, & \mathcal{H}_{1}, \end{cases} \\
\label{Eq-Eve-Power2}
\end{equation}
where 
$ \Delta_1 
= P_0 \eta (d_{p_\text{T}}^e )^{-2}$, and
$ \Delta_2 
=  P_0 \eta (d_{p_\text{T}}^e)^{-2} + \zeta \kappa P_0 \eta^2 |g_b^e|^2 ( d_{p_\text{T}}^b )^{-2} (d_b^e )^{-\alpha}$. Since the noise power at $e$ cannot be precisely obtained in the dynamic environment, we consider a bounded noise uncertainty model, where the exact noise power $\sigma_e^2$ lies within a finite range around a nominal noise power $\widetilde{\sigma}_e^2$ \cite{wu2025covert}. We assume that the dB domain $\sigma_{e, \text{dB}}^2 \in [\widetilde{\sigma}_{e, \text{dB}}^2 - \varrho_{\text{dB}}, \widetilde{\sigma}_{e, \text{dB}}^2 + \varrho_{\text{dB}}]$ follows a uniform distribution. Then the probability density function (PDF) of $\sigma_{e, \text{dB}}^2$ is $f_{\sigma_{e, \text{dB}}^2}(y) = 1/(2\rho_{\text{dB}})$, where $\sigma_{e, \text{dB}}^2 = 10\log_{10}(\sigma_e^2)$, $\widetilde{\sigma}_{e, \text{dB}}^2 = 10\log_{10}(\widetilde{\sigma}_e^2)$, and $\varrho_{\text{dB}}=10\log_{10}(\varrho)$, $\varrho > 1$.
The value of $\varrho$ measures the level of noise uncertainty \cite{wu2025covert}. The PDF of $\sigma_e^2$ as $f_{\sigma_e^2}\left(x\right) = 1/(2 x \ln \left( \varrho \right)), \ x \in \left[\check{x}, \hat{x}\right]$, 
where $\check{x} = \widetilde{\sigma}_e^2/\varrho$ and $\hat{x} = \varrho \widetilde{\sigma}_e^2$. 

Recall that Eve aims to makes the optimal decision for binary hypothesis testing and minimizes the DEP. The detection errors of $e$ consists of two types, i.e., the false alarm and the miss detection. The former denotes that $e$ makes a $\mathcal{D}_1$ decision while $b$ does not transmit (i.e., $\mathcal{H}_0$), the latter is that $e$ makes a $\mathcal{D}_0$ decision while $b$ is transmitting (i.e., $\mathcal{H}_1$). Hence, the total DEP can be calculated by $\mathcal{P}_{\text{total}} \triangleq \mathcal{P}_f + \mathcal{P}_m$~\cite{wu2025covert}, 
where $\mathcal{P}_f$ and $\mathcal{P}_m$ denote the false alarm and miss detection probabilities, respectively. Their values can be calculated by
\begin{equation}
\small
\mathcal{P}_f \triangleq \mathbb{P}\left(\mathcal{D}_1|\mathcal{H}_0\right) 
= \mathbb{P}\left(\Delta_1 + \sigma_e^2 \geq \Gamma_{\text{th}} \right).
\label{Eq-Pf}
\end{equation}
\begin{equation}
\mathcal{P}_m \triangleq \mathbb{P}(\mathcal{D}_0|\mathcal{H}_1)
= \mathbb{P}\left(\Delta_2 + \sigma_e^2 \leq \Gamma_{\text{th}} \right).
\label{Eq-Pm}
\end{equation}
The formulae of $\mathcal{P}_f,\mathcal{P}_m$ and $\mathcal{P}_{\text{total}}$ are derived in Appendix~~\ref{appx}. Based on the the monotonicity of $\mathcal{P}_{\text{total}}$, we can obtain the optimal detection threshold $\Gamma_{\text{th}}^*$ at $e$ along with the corresponding minimum total DEP $\mathcal{P}_{\text{total}}^*$ of $e$, which is given in~\cref{lemma1}.
\begin{lemma}\label{lemma1}
The optimal detection threshold minimizing the total error probability at Eve is $\Gamma_{\text{th}}^* = \check{x} + \Delta_2$, and the corresponding minimum total DEP is given by 
$$
\mathcal{P}_{\text{total}}^* = \frac{1}{2\ln \left( \varrho \right)} \ln \left(\frac{ \hat{x} } {\check{x} + \kappa P_0 \left|h_{f_\text{T}}^b \right|^2  \left|h_b^e\right|^2}\right) 
.$$
\end{lemma}

\vspace{-0.3cm}
\subsection{Detection Performance From RPA’s Perspective}
Since Eve may conceal and change its location, it is difficult for $p_{\mathrm{R}}$ to acquire its location and the perfect channel state information (CSI) of the BD–to-Eve link~\cite{wu2025covert}. 
As such, we consider a practical scenario where only imperfect estimates of Eve’s location and CSI of the BD-to-Eve link are available at $p_{\mathrm{R}}$. For the location uncertainty of $e$, we adopt a bounded location estimation error model $\mathbf{e} \in \Theta \triangleq \{ \|\mathbf{e} - \tilde{\mathbf{e}}\| \leq \chi \}$, 
where $\tilde{\mathbf{e}}$ denotes the estimated location of $e$ and $\chi$ is the maximum location estimation error. For the CSI uncertainty of the BD–to-Eve link, we consider a bounded CSI estimation error model $|g_b^e - \tilde{g}_b^e| \leq \delta$,
where $\tilde{g}_b^e $ represents the estimated CSI of BD-Eve link and $\delta$ denotes the maximum CSI estimation error.
When both the location and CSI estimations are at their maximum errors (i.e., $\hat{l}_b^e$ and $\hat{g}_b^e$), the scenario represents the worst case, resulting in the minimum detection error lower bound $\check{\mathcal{P}}_{\text{total}}^*$, which is expressed as
\begin{equation}
\small
\check{\mathcal{P}}_{\text{total}}^* = \frac{1}{2 \ln \left( \varrho \right)} \ln \left( \frac{ \hat{x} }{\check{x} + \kappa P_0 |h_{f_\text{T}}^b|^2 |\hat{h}_b^e |^2 } \right),
\label{Eq-Pro-total-error}
\end{equation}
where $\hat{h}_b^e = \hat{l}_b^e \hat{g}_b^e$, $\hat{g}_b^e = \left(1 + \delta\right) \tilde{g}_b^e$, $
\hat{l}_b^e = \max_{\mathbf{e} \in \Theta} h_b^e = \sqrt{\eta (\|\mathbf{b} - \tilde{\mathbf{e}} \| - \chi )^{-\alpha}}$, 
and $\mathbf{e} = \tilde{\mathbf{e}} + \frac{\mathbf{b} - \tilde{\mathbf{e}}}{\| \mathbf{b} - \tilde{\mathbf{e}} \|} \chi$ is the closest boundary point along the BD–to-Eve direction $\Theta$.


\vspace{-5pt}
\section{Problem Formulation and Solution}\label{Sec-Problem}
This section first formulates the optimization problem of the transmit power and the position of TPA and then solve this problem using an alternating optimization (AO) approach.

\vspace{-5pt}
\subsection{Problem Formulation}
We aim to maximize the covert transmission rate by jointly optimizing the transmit power and antenna positions while satisfying both communication reliability and covertness constraints.  
The optimization problem is formulated as follows:
\vspace{-0.6cm}

{
\begin{subequations}\label{Eq-Goal-1}
\small
    \begin{align}  
        \textbf{P1}: \max_{P_0, x_p^t} & \ R_b \label{Eq-Goal-main-1} \\
        \text{s.t. }
        & \gamma_b \geq \gamma_{\text{th}} \label{Eq-C2-1}, \\
        & \check{\mathcal{P}}_{\text{total}}^* \geq 1 - \epsilon ,\label{Eq-C3-1} \\
        & 0 \leq P_0 \leq \hat{P}, \label{Eq-C4-1} \\
        & 0 \leq x_p^t \leq L. \label{Eq-C5-1}
    \end{align}
\end{subequations}
}%
In \textbf{P1}, $\hat{P}$ is the maximum transmit power of $p_{\mathrm{T}}$.
~\cref{Eq-C2-1} ensures successful communication by requiring the SNR at $p_{\mathrm{R}}$ to exceed the threshold $\gamma_{\text{th}}$.~\cref{Eq-C3-1} guarantees the covertness of the backscatter transmission by requiring the total DEP $\check{\mathcal{P}}_{\text{total}}^*$ to be no less than $1 - \epsilon$, where $\epsilon \in [0, 1]$ is typically set to a small value to ensure covertness.~\cref{Eq-C4-1} ensures that the transmit power does not exceed $\hat{P}$.~\cref{Eq-C5-1} restricts the position of $p_{\mathrm{T}}$ along the waveguide to lie within the feasible interval $[0, L]$. The problem in~\cref{Eq-Goal-1} is non-convex due to the coupling between $P_0$ and $x_p^t$ of $p_{\mathrm{T}}$ in both the objective function and constraints. To solve it efficiently, we adopt the AO approach to decompose $\textbf{P1}$ into \textbf{transmit power optimization} subproblem and \textbf{TPA position optimization} subproblem solving $P_0$ and $x_p^t$ in an iterative manner.

\vspace{-5pt}
\subsection{Transmit Power Optimization Subproblem}
Given a fixed $x_p^t$, $P_0$ is optimized through the following subproblem:
\vspace{-0.6cm}

{
\begin{subequations}\label{Eq-Goal-2}
\small
    \begin{align}  
        \textbf{P2}: \max_{P_0} & \ \gamma_b \label{Eq-Goal-main-2} \\
        \text{s.t. }
        & P_0 \geq \check{P}_{\text{SNR}},  \label{Eq-C2-2} \\
        & P_0 \leq \hat{P}_{\text{Covert}}, \label{Eq-C3-2} \\
        & ~\cref{Eq-C4-1}.
    \end{align}
\end{subequations}
}%
In \textbf{P2}, $\check{P}_{\text{SNR}} = \gamma_{\text{th}}\kappa^{-1} (\sigma_p d_{p_\text{T}}^b d_b^{p_\text{R}}\eta^{-1} )^2$ and 
$\hat{P}_{\text{Covert}} = \xi (d_{p_\text{T}}^b )^2 (\kappa \eta |\hat{h}_b^e |^2)^{-1}$, where $\xi = \hat{x} \varrho^{-2 \left(1 - \epsilon \right) } - \check{x} $.
Hence, we can obtain $P_0 \in [\max(0, \check{P}_{\text{SNR}} ), \min(\hat{P}, \hat{P}_{\text{Covert}} )]$ when $\check{P}_{\text{SNR}}\leq$ $ \min(\hat{P}, \hat{P}_{\text{Covert}} )$.
Recall that $\gamma_b$ is monotonically increasing with respect of $P_0$. Hence, the optimal transmit power $P_0^*$ is given by the following~\cref{Theo-Power}.

\begin{theorem}\label{Theo-Power}
The optimal transmit power $P_0^*$ is given by
\begin{equation*}
    P_0^* = \min\left(\hat{P}, \hat{P}_{\text{Covert}} \right),
\end{equation*}
when $\check{P}_{\text{SNR}} \leq \min\left(\hat{P}, \hat{P}_{\text{Covert}} \right)$.
\end{theorem}

\vspace{-5pt}
\subsection{TPA Position Optimization Subproblem} 
Given a fixed $P_0$, $x_p^t$ can be optimized by maximizing $\check{P}_{\text{SNR}}$, which is equivalent to minimizing $\left(d_{p_\text{T}}^b d_b^{p_\text{R}}\right)$, where
$$d_{p_\text{T}}^b = \sqrt{(x_p^t - x_b)^2 + (y_w^t - y_b)^2 + H^2},$$
$$d_b^{p_\text{R}} = \sqrt{((x_p^r - x_b)^2 + (y_p^r - y_b)^2 + H^2}.$$ Since $d_b^{p_\text{R}}$ is independent of $x_p^t$, $\left(d_{p_\text{T}}^b d_b^{p_\text{R}}\right)$ is minimized by minimizing $d_{p_\text{T}}^b$. Thus, $x_p^t$ is optimized through the following subproblem:
\vspace{-0.3cm}

{
\begin{subequations}\label{Eq-Goal-3}
\small
    \begin{align}  
        \textbf{P3}: \min_{x_p^t} &   \left(x_p^t - x_b\right)^2 \label{Eq-Goal-main-4} \\
        \text{s.t. }
        &  x_p^t \in \left[\check{r}_{\text{SNR}}, \hat{r}_{\text{SNR}} \right],  \label{Eq-C2-4} \\
        & x_p^t \leq \check{r}_{\text{Covert}}, \ x_p^t \geq \hat{r}_{\text{Covert}}, \label{Eq-C3-4} \\
        & ~\cref{Eq-C5-1},
    \end{align}
\end{subequations}
}%
where 
\vspace{-0.3cm}

{\footnotesize
\begin{align*}
    &\check{r}_{\text{SNR}} = x_b -  \sqrt{ \frac{\kappa P_0 \eta^2}{\gamma_{\text{th}} \sigma_p^2  \left(d_b^{p_\text{R}}\right)^2} - \Delta_3 },\check{r}_{\text{Covert}} = x_b - \sqrt{\frac{\kappa \eta P_0 |\hat{h}_b^e|^2}{\xi} - \Delta_3 },\\
    &\hat{r}_{\text{SNR}} = x_b +  \sqrt{\frac{\kappa P_0 \eta^2}{\gamma_{\text{th}} \sigma_p^2  \left(d_b^{p_\text{R}}\right)^2} - \Delta_3 }, \hat{r}_{\text{Covert}} = x_b + \sqrt{\frac{\kappa \eta P_0 |\hat{h}_b^e|^2}{\xi} - \Delta_3 },
\end{align*}
}%
and $\Delta_3 = \left(y_w^t + y_b\right)^2 - H^2$. Solve \textbf{P3}, the optimal position $x_p^{t*}$ is given in~\cref{Theo-Position}.
\begin{theorem}\label{Theo-Position}
The optimal TPA position is given by
\begin{equation*}
    x_p^{t*} = \begin{cases}
    \check{r}_{\text{Covert}}, & \check{r}_{\text{Covert}} \geq  \max[0, \check{r}_{\text{SNR}}],  \\
    \hat{r}_{\text{Covert}},   & \hat{r}_{\text{Covert}} \leq  \min[L, \hat{r}_{\text{SNR}}]. \\
\end{cases}
\end{equation*}
\end{theorem}

\section{Numerical Analysis}\label{Sec-Ana}
In this section, numerical results are provided to demonstrate the effectiveness of our proposed algorithm. For simplicity, we consider a square room $\mathcal{A}$ with $L=D=20$m. The BD is positioned at the center of $\mathcal{A}$ with $\mathbf{b}: \left(L/2, 0, 0\right)$, and the EV is located at a distance $d_b^e$ from $b$. The RPA is fixed at the center of $w_{\mathrm{R}}$ with $x_p^r = L/2$\footnote{It is feasible to extend the framework to jointly optimize both TPA and RPA, but would introduce more complex non-convex coupling between the downlink energy harvesting and uplink backscatter links without yielding additional structural insights, and is therefore left for future work.}. In our results, the default parameter configurations are detailed as follows: $n_{\rm eff}=1.4$, $f_c=28$GHz~\cite{ding25flexible}, $\kappa=0.375$, $\zeta = 1$, $\alpha=2$, $B=10$kHz, $\gamma_{\text{th}} = 0$dB, $\tilde{g}_b^e = 1.278$, and $\varrho_{\text{dB}} = 3$dB~\cite{wu2025covert}. Besides, the values of other parameters are specified in the captions of the respective figures. To provide a performance benchmark, we also consider a baseline algorithm in which the TPA position is fixed at $x_p^t = 0$, $L/4$, and $L/2$, and then $P_0$ is optimized. This baseline imitates a \textbf{conventional fixed-position antenna system}\footnote{Since the BD is located at the center of $\mathcal{A}$, the cases $x_p^t = 3L/4$ and $L$ are symmetric to $x_p^t = 0$ and $L/4$, respectively.}. In each figure, the red curves represent our results while the blue ones represent baseline's.

\begin{figure}[t]
    \captionsetup[subfigure]{justification=centering}
    \centering
    \subfloat[$\hat{R}_b$ versus $d_b^e$, $\chi$ and $\delta$.]{\includegraphics[width=6 cm]{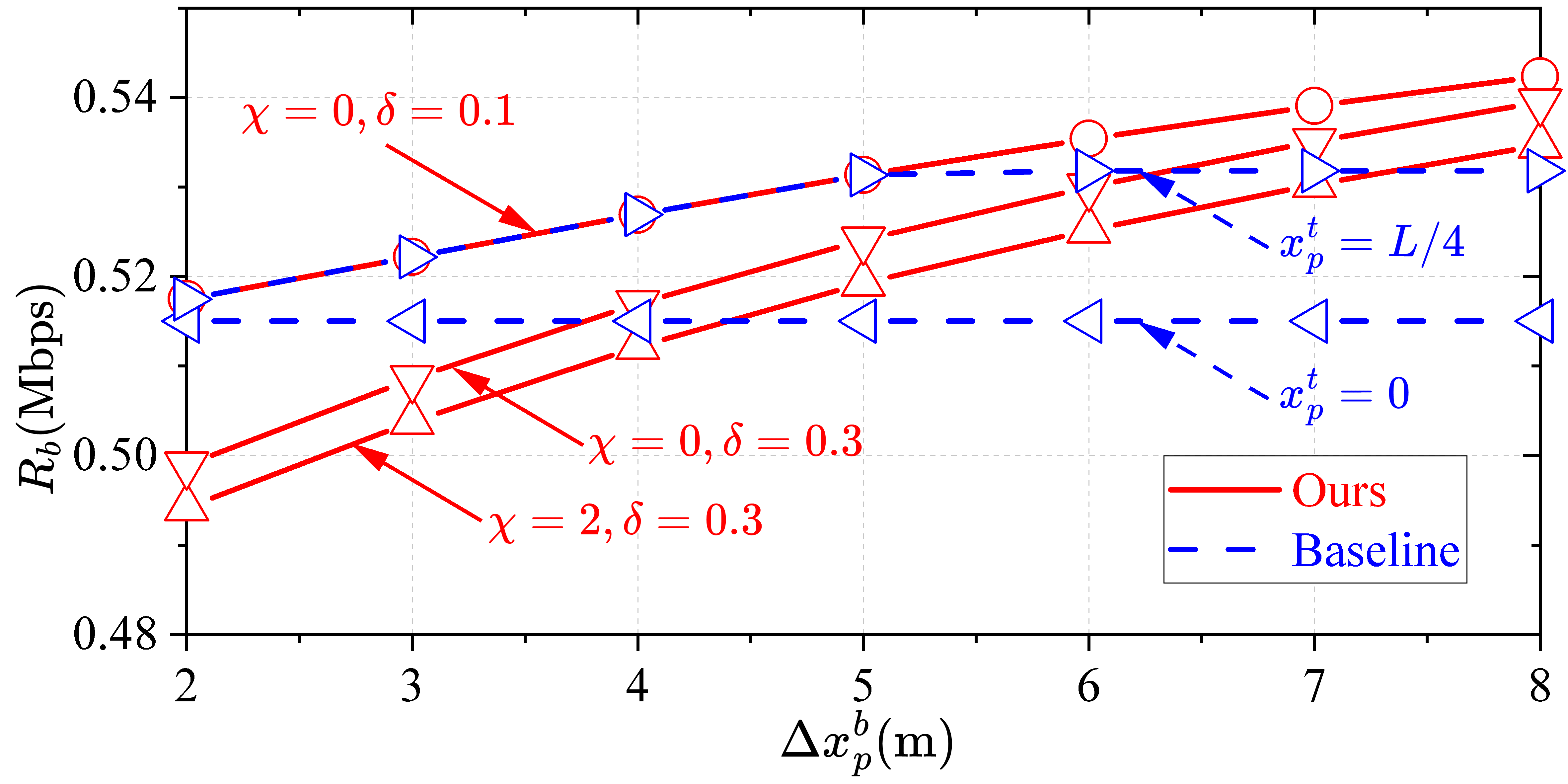}\label{Fig-Opt-D-be-Rate}}	\hfil 
    \subfloat[$\hat{R}_b$ versus $P_0$ and $\widetilde{\sigma}_e^2$.]{\includegraphics[width=6 cm]{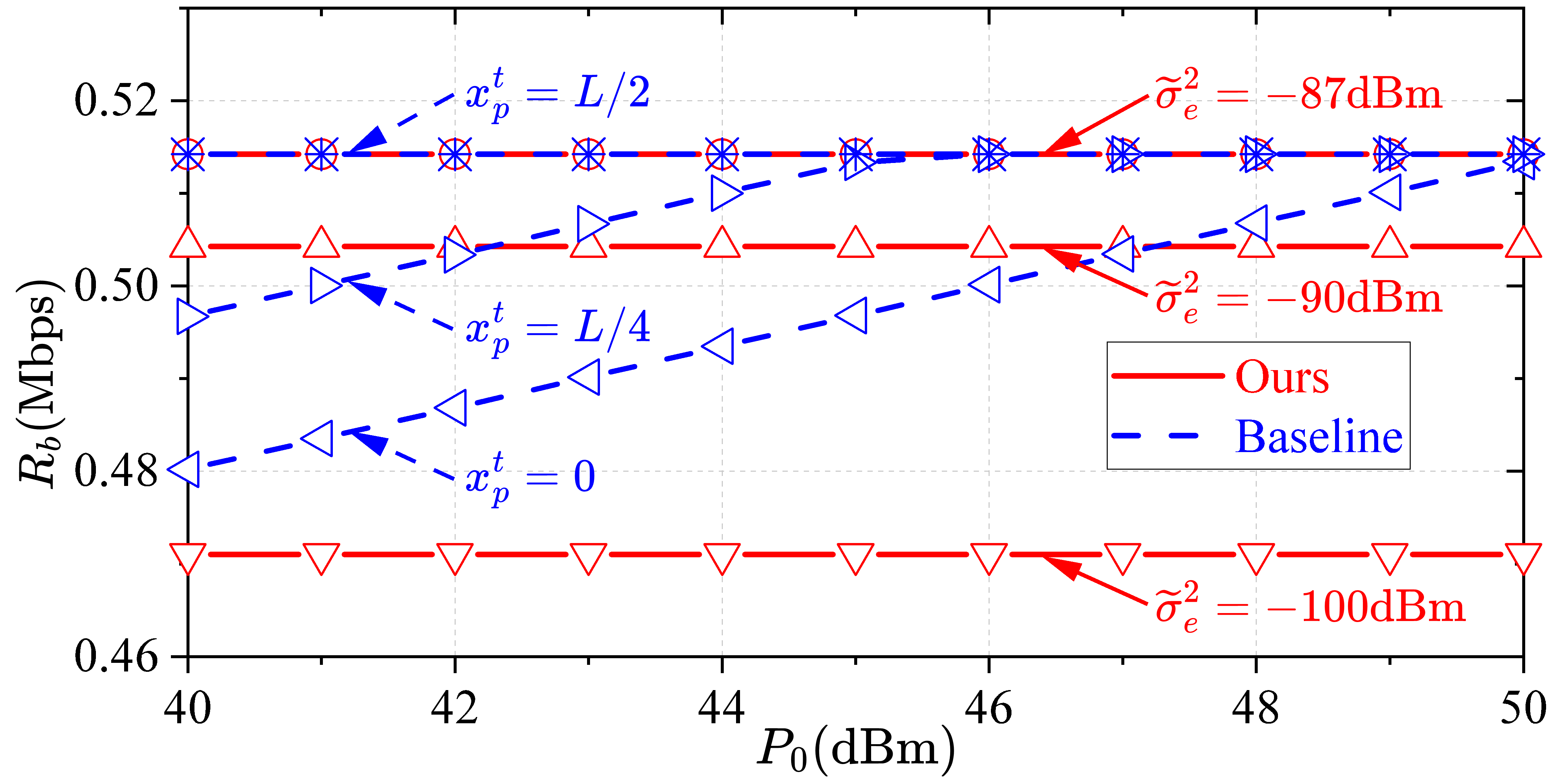}\label{Fig-Opt-Power-Rate}}	\hfil 
    \subfloat[$\hat{R}_b$ versus $\epsilon$ and $\sigma_p^2$.]{\includegraphics[width=6 cm]{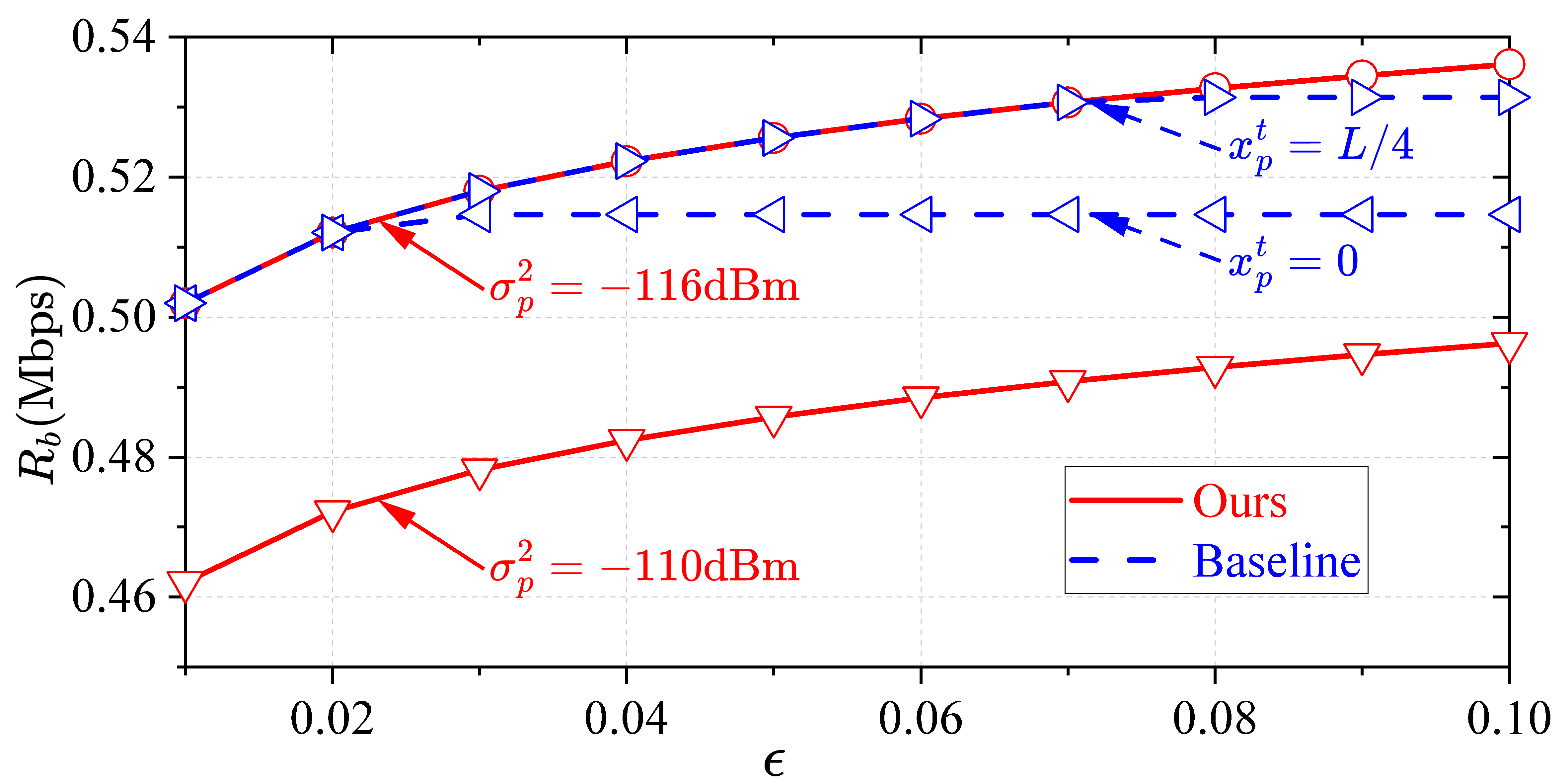}\label{Fig-Opt-Epsilon-Rate}}	\hfil 
\caption{$\hat{R}_b$ versus various parameters. Specifically, in (a) $\chi = \{0, 2\}$m, $\delta = \{0.1, 0.3\}$, $P_0=50$dBm, $\sigma_p^2 = -116$dBm, $\widetilde{\sigma}_e^2 = -90$dBm, $\epsilon = 0.05$, $y_w^t = -0.5$m, and $y_w^r = 0.5$m; in (b), $\widetilde{\sigma}_e^2 = \{-87, -90, -95\}$dBm, $\sigma_p^2 = -116$dBm, $d_b^e = 5$m, $\chi = 2$m, $\delta = 0.3$, $\epsilon = 0.02$, $y_w^t = -1$m, and $y_w^r = 1$m; in (c), $\sigma_p^2 = \{-110, -116\}$dBm, $P_0=50$dBm, $\widetilde{\sigma}_e^2 = -90$dBm, $d_b^e = 5$m, $\chi = 1$m, $\delta = 0.3$, $y_w^t = -1$m, and $y_w^r = 1$m.}
\label{Fig-Ps-parameter}
\vspace{-0.1cm}
\end{figure}

In~\cref{Fig-Opt-D-be-Rate}, we analyze the impacts of $d_b^e$, $\chi$ and $\delta$ on $\hat{R}_b$. It can be observed that given $\chi$ and $\delta$, $\hat{R}_b$ increases as $d_b^e$ increases. This is because as $d_b^e$ increases, Eve is more likely to be positioned further from the BD, hence, the TPA can enhance its transmit power $P_0$, which boosts the backscattered power of $b$ and the received power at $p_{\mathrm{T}}$, thereby increasing $\hat{R}_b$. Besides, given $d_b^e$ and $\delta$, an increase in $\chi$ leads to a decrease in $\hat{R}_b$. Similarly, given $d_b^e$ and $\chi$, an increase in $\delta$ also results in a decrease in $\hat{R}_b$. According to~(\ref{Eq-Pro-total-error}), larger values of $\chi$ and $\delta$ correspond to a larger $\hat{h}_b^e$, which leads to a lower minimum detection error. As a result, the TPA reduces $P_0$, thereby decreasing $\hat{R}_b$. In addition, we compare with the baseline algorithm under the setting $\chi = 0$ m and $\delta=0.1$. It can be observed that when $x_p^t = 0$, $\hat{R}_b$ remains constant and is lower than that in our algorithm. This occurs because that TPA is too far from the BD, transmitting with $P_0^* = \hat{P}_0$; when $x_p^t = L/4$, $\hat{R}_b$ initially increases and reaches the value seen in our algorithm, then remains constant which is lower that in ours. This behavior is due to the increase in $P_0^*$ until it reaches $P_0^* = \hat{P}_0$; and when $x_p^t = L/2$, $\hat{R}_b=0$. It is because the TPA is too close to the BD, violating the covertness constraint of the BD for any $P_0^*$.

In~\cref{Fig-Opt-Power-Rate}, we analyze the impacts of $P_0$ and $\widetilde{\sigma}_e^2$ on $\hat{R}_b$. It can be observed that given $\widetilde{\sigma}_e^2$, $\hat{R}_b$ remains unchanged as $P_0$ increases. This is because the distance $d_b^e$, the channel conditions (e.g., $\chi$, $\delta$), and minimum error probability threshold are fixed, meaning that $\hat{R}_b$ is predetermined. Given $P_0$, increasing $\widetilde{\sigma}_e^2$ leads to a larger $\hat{R}_b$. This is because a larger value of $\widetilde{\sigma}_e^2$ yields a higher minimum error probability, allowing the TPA to improve $P_0$, which enhances $\hat{R}_b$.
Additionally, we compare with the baseline algorithm under the setting $\widetilde{\sigma}_e^2 = -87$ dBm. It can be observed that when $x_p^t = 0$, $\hat{R}_b$ gradually increases but remains lower than that in ours. This is because the TPA is too far from the BD, transmitting with $P_0^* = \hat{P}_0$; when $x_p^t = L/4$, $\hat{R}_b$ initially increases it matches the value in ours, then remains constant. This behavior is due to the increase in $P_0^*$ until reaches to $P_0^* = \hat{P}_0$; and when $x_p^t = L/2$, $\hat{R}_b$ always equals to that in ours. This is because the settings ensure that the transmission always satisfies the covertness constraint of BD, allowing the TPA to be positioned at $L/2$.

In~\cref{Fig-Opt-Epsilon-Rate}, we analyze the impacts of $\epsilon$ and $\sigma_p^2$ on $\hat{R}_b$. It can be observed that given $\sigma_p^2$, $\hat{R}_b$ increases as $\epsilon$ increases. This is because a larger $\epsilon$ (i.e., a smaller $1 - \epsilon$) corresponds to a lower threshold for the minimum error probability. Therefore, the TPA can improve $P_0$, leading to an increase in $\hat{R}_b$. Given $\epsilon$, $\hat{R}_b$ decreases as $\sigma_p^2$ increases. It occurs because a larger $\sigma_p^2$ improves the decoding threshold at $p_{\mathrm{R}}$, thereby reducing $\hat{R}_b$.
Besides, we compare with the baseline algorithm under the setting $\sigma_p^2 = -116$ dBm. It can be observed when $x_p^t = 0$ and $L/4$, $\hat{R}_b$ initially increases, reaching the value observed in ours, and then remains constant which is lower that in ours. However, the maximum value for $x_p^t = L/4$ is higher than that for $x_p^t = 0$. This behavior is due to the increase in $P_0^*$ until it reaches $P_0^* = \hat{P}_0$; and when $x_p^t = L/2$, $\hat{R}_b$ is zero. This is because the TPA is too close to the BD, violating the covertness constraint of the BD.

\vspace{-5pt}
\section{Conclusion}\label{Sec-Con}
This paper investigated the uplink covert transmission rate maximization in a PA-BSC network, where the BD simultaneously harvests energy from the TPA and backscatters data to the RPA while being subject to detection by a randomly located Eve. A 3D model was developed to characterize the spatial distributions of the BD, Eve, and PAs, along with a covert communication model that captures Eve’s detection capabilities. We formulated and solved an optimization problem by alternately optimizing the TPA’s transmit power and location. The simulation results demonstrate the effectiveness of the proposed algorithm, showing that the system mitigates the double near-far effect and achieves significant improvements in system performance compared to baseline scheme.

\vspace{-5pt}
{\appendices

{\appendices
\section{}
\label{appx}
\vspace{-0.3em}
The probabilities $\mathcal{P}_f$ and $\mathcal{P}_m$ are respectively calculated by
\begin{equation}
\small
\mathcal{P}_f = \begin{cases}1, & \Gamma_{\text{th}} < \check{x} + \Delta_1, \\ 
\theta_{1}, & \check{x} + \Delta_1 \leq \Gamma_{\text{th}} \leq \hat{x} + \Delta_1, \\
0, & \Gamma_{\text{th}} > \varrho \hat{x} + \Delta_1, \end{cases}
\label{Eq-Pf}
\end{equation}
\begin{equation}
\mathcal{P}_m = \begin{cases}0, & \Gamma_{\text{th}} < \check{x} + \Delta_2, \\ 
\theta_2, & \check{x} + \Delta_2 \leq \Gamma_{\text{th}} \leq \hat{x} + \Delta_2, \\ 
1, & \Gamma_{\text{th}} > \hat{x} + \Delta_2, \end{cases}
\label{Eq-Pm}
\end{equation}
where 
$\theta_{1} = \int_{\Gamma_{\text{th}} - \Delta_1}^{\varrho \widetilde{\sigma}_e^2} \frac{1}{2 x \ln\left(\varrho\right)} \mathrm{d}x = \frac{1}{2\ln\left(\varrho\right)} \ln \left(\frac{\varrho \widetilde{\sigma}_e^2}{\Gamma_{\text{th}} - \Delta_1}\right)$, and
$\theta_2 = \int_{\widetilde{\sigma}_e^2/\varrho}^{\Gamma_{\text{th}} - \Delta_2}\frac{1}{2 x \ln\left(\varrho\right)}\mathrm{d}x = \frac{1}{2 \ln\left(\varrho\right)} \ln\left(\frac{\varrho(\Gamma_{\text{th}} - \Delta_2)}{\widetilde{\sigma}_e^2}\right)$.
Based on~(\ref{Eq-Pf}) and~(\ref{Eq-Pm}), by considering equal probability of $\mathcal{H}_0$ and $\mathcal{H}_1$, we derive the expression of $\mathcal{P}_{\text{total}}$ as follows:
\begin{equation}
\small
\mathcal{P}_{\text{total}} = \begin{cases}1, & \Gamma_{\text{th}} < \check{x} + \Delta_1, \\ 
\theta_{1}, & \check{x} + \Delta_1 \leq \Gamma_{\text{th}} < \check{x} + \Delta_2, \\ 
\theta_{1} + \theta_2, & \check{x} + \Delta_2 \leq \Gamma_{\text{th}} \leq \hat{x} + \Delta_1, \\ 
\theta_2, & \hat{x} + \Delta_1 < \Gamma_{\text{th}} \leq \hat{x} + \Delta_2, \\ 
1, & \Gamma_{\text{th}} > \hat{x} + \Delta_2.\end{cases}
\label{Eq-Ptotal-detail}
\end{equation}

\noindent From (\ref{Eq-Ptotal-detail}), we can obtain the monotonicity of $\mathcal{P}_{\text{total}}$: 1) when $\check{x} + \Delta_1 \leq \Gamma_{\text{th}} < \check{x} + \Delta_2$, $\mathcal{P}_{\text{total}}=$ $\frac{1}{2\ln\left(\varrho\right)} \ln \left(\frac{\varrho \widetilde{\sigma}_e^2}{\Gamma_{\text{th}} - \Delta_1}\right)$, which is monotonically decreasing with $\Gamma_{\text{th}}$; 2) when $\check{x} + \Delta_2 \leq \Gamma_{\text{th}} \leq \hat{x} + \Delta_1$, $\mathcal{P}_{\text{total}} = \frac{1}{2\ln \left( \varrho \right)} \ln \left( \varrho^2 - \frac{\varrho^2 \kappa P_0 |h_{f_\text{T}}^{p_\text{T}}|^2 |h_{p_\text{T}}^b|^2 |h_b^e|^2}{\Gamma_{\text{th}} - P_0 |h_{f_\text{T}}^{p_\text{T}}|^2 |h_1^{\mathrm{e}}|^2}\right)$, which is monotonically increasing with $\Gamma_{\text{th}}$; 
3) when $\hat{x} + \Delta_1 < \Gamma_{\text{th}} \leq \hat{x} + \Delta_2$, $\mathcal{P}_{\text{total}} = \frac{1}{2 \ln\left(\varrho\right)} \ln\left(\frac{\varrho(\Gamma_{\text{th}} - \Delta_2)}{\widetilde{\sigma}_e^2}\right)$, which is monotonically increasing with $\Gamma_{\text{th}}$.
}
}
\vspace{-0.3cm}

\bibliographystyle{IEEEtran}
\bibliography{IEEEabrv, references.bib}

\end{document}